\documentclass[a4paper,12pt]{article}
\usepackage{amsfonts}
\usepackage{amssymb,latexsym}
\makeatletter \@addtoreset{equation}{section} \makeatother

\mathchardef\varGamma="0100 \mathchardef\varDelta="0101
\mathchardef\varTheta="0102 \mathchardef\varLambda="0103
\mathchardef\varXi="0104 \mathchardef\varPi="0105
\mathchardef\varSigma="0106 \mathchardef\varUpsilon="0107
\mathchardef\varPhi="0108 \mathchardef\varPsi="0109
\mathchardef\varOmega="010A

\def\bfone{\relax{\rm 1\kern-.35em 1}}

\DeclareFontFamily{U}{rsf}{} \DeclareFontShape{U}{rsf}{m}{n}{
  <5> <6> rsfs5 <7> <8> <9> rsfs7 <10-> rsfs10}{}
\DeclareMathAlphabet\Scr{U}{rsf}{m}{n}

\newcommand{\rU}{\mathrm{U}}
\newcommand{\rUSp}{\mathrm{USp}}
\newcommand{\rSU}{\mathrm{SU}}
\newcommand{\rE}{\mathrm{E}}
\newcommand{\rSO}{\mathrm{SO}}
\newcommand{\rO}{\mathrm{O}}
\newcommand{\rSL}{\mathrm{SL}}

\newcommand{\rGL}{\mathrm{GL}}

\begin{document}

\begin{titlepage}

\thispagestyle{empty}

\begin{flushright}
\hfill{CERN-PH-TH/135}
\end{flushright}

\vspace{35pt}

\begin{center}{ \LARGE{\bf
Curvatures and potential of M--theory in\\[5mm] $D=4$
with fluxes and twist}} \vspace{60pt}

{\bf  R. D'Auria$^\bigstar$, S. Ferrara$^\dag$ and M.
Trigiante$^\bigstar $}

\vspace{15pt}

$^\bigstar${\it Dipartimento di Fisica, Politecnico di Torino \\
C.so Duca degli Abruzzi, 24, I-10129 Torino, and\\
Istituto Nazionale di Fisica Nucleare, \\
Sezione di Torino,
Italy}\\[1mm] {E-mail: riccardo.dauria@polito.it,  mario.trigiante@polito.it}

$^\dag$ {\it CERN, Physics Department, CH 1211 Geneva 23,
Switzerland\\ and\\ INFN, Laboratori
Nucleari di Frascati, Italy\\and\\
Department of Physics \& Astronomy, University of California, Los Angeles, CA, USA}\\[1mm] {E-mail: Sergio.Ferrara@cern.ch}

\vspace{50pt}

{ABSTRACT}

\end{center}

\medskip

We give the curvatures of the  free differential algebra (FDA) of
M--theory compactified to $D=4$ on a twisted seven--torus with the
4--form flux switched on. Two formulations are given, depending on
whether the 1--form field strengths of the scalar fields
(originating from the 3--form gauge field $\hat{A}^{(3)}$) are
included or not in the FDA. We also give the bosonic equations of
motion and discuss at length the scalar potential which  emerges
in this type of compactifications. For flat groups we show the
equivalence of this potential with a dual formulation of the
theory which has the full $\rE_{7(7)}$ symmetry.

\end{titlepage}

\newpage
\section{Introduction}
Higher dimensional theories of supergravity include $(p+1)$--form
gauge fields which may couple to $p$--branes, as well as
$(D-3-p)$--form magnetic potentials which have ``magnetic''
$(D-4-p)$--branes as sources (for a review see \cite{dkl}). When these theories are compactified
to lower--dimensions, $S$ and $T$--dualities, as well as extended
supersymmetry, give rise to non--compact global symmetries \cite{cj} called
U--dualities \cite{ht}. Of particular interest is the case in which the
manifold of compactification is a ``generalized'' Calabi--Yau
manifold (in which case the $\rSU(N)$ holonomy is replaced by an
$\rSU(N)$ structure); a variety of fluxes are introduced
 and these result in ``massive'' deformations of the original
theory without fluxes \cite{ccdlmz}--\cite{glw}.\par In the low--energy effective supergravity
theory these massive deformations can, in most cases, be interpreted as gauged supergravities,
in which the non--trivial gauge group (i.e. the gauge symmetry under which some of the
$p$--forms are ``charged'') is a remnant of the ``continuous'' U--duality symmetry of the
effective supergravity theory (for a review see \cite{adfl_proc,dft_proc}).\par However,
although this is the correct description when the bosonic sector of the theory contains only
spin 1 and 0 fields (other than the graviton), the gauge structure exhibit a more general form
than that of an ordinary Lie algebra, in the case in which the gauge potentials of rank higher
than one occur in the theory. This generalized gauge structure is called free differential
algebra (FDA) \cite{fd,fda,aipv}.
\section{M--theory FDA}
The prototype of this phenomenon is the gauge structure of
eleven--dimensional supergravity \cite{cjs} compactified on a twisted seven--torus with a 4--form
flux turned on \cite{ss}-\cite{dft2}. This is a ``massive'' deformation of the original
$D=11$ supergravity theory compactified on $T^7$, considered by
Cremmer and Julia in their derivation of the $N=8,\,D=4$
supergravity. However the deformation occurs before seven
antisymmetric tensor fields $B_{\mu\nu I}$ have been dualized into
the scalars $\tilde{B}^I$ and so the gauge structure is not encoded in a Lie algebra
which is a subalgebra of $\rE_{7(7)}$.\par The resolution of this
puzzle was found in \cite{ddf,dft2} by noting that the 28 gauge potentials
$A^I_\mu=g^I_\mu$, $A_{\mu IJ}$ and the seven antisymmetric
tensors $B_{\mu\nu I}$ have the  structure of a free differential
algebra where the curvature 2--forms of the (would be) Lie algebra
part are\footnote{In the present work we have changed the normalizations of $g_{IJKL}$ and $B_I$ with respect to \cite{dft2}, so that the following formulae can be obtained from the corresponding ones in \cite{dft2} through the substitution: $g\rightarrow -12\,g$ and $B_I\rightarrow -B_I/2$.}:
\begin{eqnarray}
F^I &=& dA^I + \frac{1}{2}\,\tau_{JK}{}^I\, A^J \wedge A^K \,,\label{eq1}\\
 F^{(0)}_{IJKL}&=&- g_{IJKL}\,,\label{eq2}\\
{\Scr F}_{IJ} &=& {\Scr D}^{(\tau)}A_{IJ}-6\,
 g_{IJKL}\,A^K \wedge A^L-\frac{1}{2}\,\tau_{IJ}{}^K \, B_K \,,\label{eq0}
\end{eqnarray}
where the connection of the ${\Scr D}^{(\tau)}$ covariant
derivative is:
\begin{eqnarray}
\omega^I{}_J&=&-\tau_{JK}{}^I\,A^K\,\,\,;\,\,\,\,\omega_I{}^J=\tau_{IK}{}^J\,A^K\,,\label{eq4}
\end{eqnarray}
and $g_{IJKL}$ is the non trivial flux in internal space.\\
 Since $B_K$ enter the expression
of ${\Scr F}_{IJ}$, this is not an ordinary Lie algebra. Moreover the $B_I$ curvature is:
\begin{eqnarray}
H_I&=&{\Scr D}^{(\tau)}B_I+4\,g_{IJKL}\,A^J\wedge A^K\wedge A^L+2\, F^J\wedge
A_{JI}\,,\label{eq5}
\end{eqnarray}
which indicates that $g_{IJKL}$ is a non--trivial cocycle of the
FDA.\par There is a further space-time 3--form ${\cal
A}^{(3)}_{\mu\nu\rho}$ whose field strength reads:
\begin{eqnarray}
F^{(4)}&=&d{\cal A}^{(3)}- g_{IJKL}\,A^I\wedge A^J\wedge A^K\wedge
A^L- B_I\wedge F^I\,.\label{eq6}
\end{eqnarray}
Equations (\ref {eq1}),(\ref {eq0}),(\ref {eq5}),(\ref {eq6}) are the 4-dimensional reduction
of the eleven dimensional equation:
\begin{equation}
\hat{F}^{(4)}= d\hat{A}^{(3)}\,,
\end{equation}
when the scalars coming from the 3-form are set to zero. Here the hats refer to the eleven
dimensional forms. Equation (\ref{eq2}) is a consequence of
the reduction of the eleven dimensional vanishing torsion.\\
We anticipate here that the above curvatures have the same structure as the ones in
\cite{dft2} with the exception of the last terms in eqs. (\ref{eq5}), (\ref{eq6}), which of
course are absent in the zero--curvature case.\\
 The main result here is that eqs.
(\ref{eq1}), (\ref{eq2}), (\ref{eq0}), (\ref{eq5}) and (\ref{eq6}) give the complete FDA for
the non--zero curvature case. The Bianchi identities involving the non--vanishing curvatures,
imply the constraints on $\tau_{IJ}{}^K,\,g_{IJKL}$ \cite{df} which were obtained as
integrability conditions in the zero--curvature case \cite{ddf} and which read:
\begin{eqnarray}
\tau_{[IJ}{}^K\,\tau_{M]K}{}^L&=&0\,,\label{eq7}\\
\tau_{[IJ}{}^K\,g_{MNL]K}&=&0\,,\label{eq8}
\end{eqnarray}
The full set of Bianchi identities is:
\begin{eqnarray}
{\Scr D}{\Scr F}_{IJ}+\frac{1}{2}\,\tau_{IJ}{}^K\,H_K&=&0\,,\label{eq9}\\
{\Scr D}F^I&=&0\,,\label{eq10}\\
{\Scr D}H_I+2\,F^J\wedge {\Scr F}_{JI}&=&0\label{eq11}\,, \\
{\Scr D}^{(\tau)}F^{(0)}_{IJKL}+6\,\tau_{IJ}{}^M\,g_{KLMN}\,A^N&=&0\,,\label{eq11.5}
\end{eqnarray}
where the covariant derivative ${\Scr D}$ acting on the field strengths is defined as follows:
\begin{eqnarray}
{\Scr D}\left(\matrix{F^I\cr F_{IJ}}\right)&=&d\left(\matrix{F^I\cr
F_{IJ}}\right)-\left(\matrix{\omega^I{}_K &0 \cr \omega_{IJ|K}&
\omega_{IJ}{}^{LN}}\right)\left(\matrix{F^K\cr F_{LN}}\right)\,,
\end{eqnarray}
the new connection being defined as:
\begin{eqnarray}
\omega^I{}_K&=&\tau_{K L}{}^I\, A^L\,,\nonumber\\\omega_{IJ|K}&=&-12\,g_{IJKL}\,
A^L+3\,\delta^{L}_{[I}\,\tau_{J]K}{}^M\,A_{LM}\,,\nonumber\\
\omega_{IJ}{}^{LN}&=&-2\,\delta^{[L}_{[I}\,\tau_{J]K}{}^{M]}\,A^K\,.
\end{eqnarray}
The Bianchi identity for $F^{(4)}$ is trivial since we are in $D=4$.\par Integrability of eqs.
(\ref{eq9}),(\ref{eq10}) and (\ref{eq11}) can be checked by further applying to them the
${\Scr D}$ operator.  As explained in reference \cite{ddf}, the set of curvatures
$(F^I,\,{\Scr F}_{IJ})$ do not form an ordinary Lie algebra since the ``structure constants''
$f_{\Lambda\Sigma}{}^\Gamma$ entering the quadratic part of:
\begin{eqnarray}
{\Scr F}^\Lambda &=&
dA^\Lambda+\frac{1}{2}\,f_{\Gamma\Sigma}{}^\Lambda\,A^\Gamma\wedge
A^\Sigma+m^{\Lambda I}\,B_I\,,\label{eq12}
\end{eqnarray}
do not satisfy the Jacobi identities of a Lie algebra:
\begin{eqnarray}
f_{[\Lambda\Sigma}{}^\Gamma\,f_{\Delta]\Gamma}{}^\Pi&\neq
&0\,,\label{eq13}
\end{eqnarray}
because of the trilinear term in the definition (\ref{eq5}) of
$H_I$.\par
Even when $g_{IJKL}=0$ the 28--dimensional Lie algebra gauged by the vectors $A^I,\,A_{IJ}$
is not a subalgebra of $\rE_{7(7)}$, as explained in \cite{dft2}. This is a consequence of the
phenomenon of ``dualization of dualities'' discussed in \cite{dd}.
\section{FDA, including scalar fields}
To make contact with the original work of ref. \cite{ss} and their definition of curvatures
(called G in \cite{ss}), it is useful to include the scalar fields:
\begin{eqnarray}
C_{IJK}&=&A_{IJK}\,,\label{eq14}
\end{eqnarray}
 and also an ``internal'' curvature 0--form :
\begin{eqnarray}
F^{(0)}_{IJKL}&=& -
g_{IJKL}-\frac{3}{2}\,\tau_{[IJ}{}^M\,C_{KL]M}\,.\label{eq15}
\end{eqnarray}
Then we may include a curvature 1--form:
\begin{eqnarray}
F^{(1)}_{IJK}&=&{\Scr
D}^{(\tau)}C_{IJK}-4\,g_{IJKL}\,A^L-\tau_{[IJ}{}^L
A_{K]L}\,,\label{eq16}
\end{eqnarray}
and a modified curvature 2--form:
\begin{eqnarray}
F^{(2)}_{IJ}&=&{\Scr F}_{IJ}-3\,C_{IJK}\,F^K\,,\label{eq17}
\end{eqnarray}
and no other modifications in $H_I^{(3)}$ and $F^{(4)}$. The new Bianchi identities are:
\begin{eqnarray}
{\Scr D}^{(\tau)}
F^{(0)}_{IJKL}&=&-\frac{3}{2}\tau_{[IJ}{}^N\,F^{(1)}_{KL]N}\,,\label{eq18}\\
{\Scr D}^{(\tau)}
F^{(1)}_{IJK}+\tau_{[IJ}{}^N\,F_{K]N}^{(2)}-4\,F^L\wedge F^{(0)}_{LIJK} &=&0\,,\label{eq19}\\
{\Scr D}^{(\tau)} F^{(2)}_{IJ}+\frac{1}{2}\,\tau_{IJ}{}^L\,H_L
+3\,F^{(1)}_{IJK}\wedge F^K &=&0\label{eq20}\\
{\Scr D}^{(\tau)} H_I+2\,F^J\wedge F^{(2)}_{JI}&=&0\,,\label{eq21}
\end{eqnarray}
(note that $F^I\wedge F^{(2)}_{IJ}= F^I\wedge {\Scr F}_{IJ}$ since
$F^I\wedge F^J\, C_{IJK}=0$). These are the Bianchi identities of
the curvatures $G^{(0)}_{IJKL},\,G^{(1)}_{\mu JKL},\,G^{(2)}_{\mu
\nu KL}\,, ,\,G^{(3)}_{\mu \nu\rho L}$ introduced in ref. \cite{ss}, with
\begin{eqnarray}
G^{(0)}&=&F^{(0)}\,\,\,;\,\,\,\,G^{(1)}=F^{(1)}\,\,\,;\,\,\,\,G^{(2)}=F^{(2)}\,\,\,;\,\,\,\,G^{(3)}=H\,.
\end{eqnarray}

\section{The equations of motion and the potential}
The bosonic equations of motion of M--theory can be obtained by
varying the Lagrangian with respect to the vielbein 1--form $V^a$
and the 3--form $\hat{A}$. \par The $g_{\mu\nu}$, $G_{IJ}$ and
$A^I$ field equations come from
 the eleven dimensional Einstein equations:
\begin{eqnarray}\label{einst}
R_{\mu\nu}-\frac{1}{2}\,g_{\mu\nu}\,R&=&T_{\mu\nu}\,,\nonumber\\
R_{\mu I}-\frac{1}{2}\,g_{\mu I}\,R&=&T_{\mu I}\,,\nonumber\\
R_{IJ}-\frac{1}{2}\,g_{IJ}\,R&=&T_{IJ}\,,
\end{eqnarray}
where $g_{\mu I}=G_{IJ}\,A^J_\mu$ and $G_{IJ}$ are the coordinates
of ${\rGL}(7)/{\rSO(7)}$. The tensor $T$ is the energy momentum
tensor of the 4--form. Incidentally we remark that in this
formulation the $R$--symmetry
 of the corresponding $N=8$ supergravity is $Spin(7)$, the eleven dimensional
gravitino gives rise to eight gravitinos which are
in the eight--dimensional spinorial representation and to spin $1/2$ which
transform in the ${\bf 8}+{\bf 48}$ of the same group.\par
 The 3--form field equations read as follows:\footnote{For the eleven dimensional equations we are using the conventions
 and notations of reference \cite {fd}}
\begin{eqnarray}
d \star F^{(4)}&=& \frac{1}{4}\, F^{(4)}\wedge  F^{(4)}\,.\label{gff}
\end{eqnarray}
Since in this paper we are mainly concerned with the general form of the scalar potential coming from the twist and the fluxes, we will carefully analyze this equation only for those entries which
receive contributions from the scalar potential. Let us write the dual of the
field equations originating from the Euler--Lagrange equations for $\mathcal{A}_{\mu\nu\rho}$ and $C_{IJK}$.  The first equation  allows us to integrate out the  $\mathcal{A}_{\mu\nu\rho}$ field in a manner which we shall
explain in a moment. This integration gives an extra contribution to the scalar potential coming from the Chern--Simons term. The second equation contains the derivative of the vacuum energy with respect to $C_{IJK}$ and contributes to the equation of motion of the $C_{IJK}$ scalar.\par
Let us define the following 4--D scalar quantity:
\begin{eqnarray}
 P&=& \frac{1}{\sqrt{-g}}\,\epsilon^{\mu_1\dots \mu_4}\, F^{(4)}_{\mu_1\dots \mu_4}\,,\label{P}
\end{eqnarray}
where $F^{(4)}_{\mu_1\dots \mu_4}$ was defined in (\ref{eq6}). For the purpose of computing the scalar potential,
 only the $d {\cal A}^{(3)}$ part of $F^{(4)}$ will be relevant.
The $\mathcal{A}_{\mu\nu\rho}$ field equation then reads:
\begin{eqnarray}
 d (V_7\,P)&=&-\frac{1}{4}\, F^{(1)}_{IJK}\, F^{(0)}_{PQRS}\,\epsilon^{IJKPQRS}\,.\label{dP}
\end{eqnarray}
For the purpose of computing the potential, only the ${\Scr D}^{(\tau)}C_{IJK}$ term in $F^{(1)}_{IJK}$ is relevant.
Equation (\ref{dP}) implies that its right hand side is a closed form. In fact
the crucial ingredient is that the term
$F^{(1)}_{IJK}\, F^{(0)}_{PQRS}\,\epsilon^{IJKPQRS}$ is an exact form on the twisted torus with fluxes, and it can be written as
\begin{eqnarray}
 F^{(1)}_{IJK}\, F^{(0)}_{PQRS}\,\epsilon^{IJKPQRS}&=&-d\left(C_{IJK}\,
 (g_{PQRS}+\frac{3}{4}\,\tau_{[PQ}^N\,C_{RS]N})\epsilon^{IJKPQRS}+\tilde{g}\right)\,,
\end{eqnarray}
where the integration constant $\tilde{g}$ \cite{d} is actually related to the dual gauge algebra in the $\rE_{7(7)}$ covariant formulation described in \cite{dft2}. From this we get the value of $V_7\, P$ to be:
\begin{eqnarray}
V_7\, P&=&\frac{1}{4}\,\left( C_{IJK}\,(g_{LPQR}+\frac{3}{4}\,\tau_{[LP}^N\,C_{QR]N})\,\epsilon^{IJKLPQR}+\tilde{g}\right)\,.\label{dP1}
\end{eqnarray}
Note the important identity:
\begin{eqnarray}
\frac{\delta P}{\delta C_{IJK}}&=&-\frac{1}{4}\,\epsilon^{IJKLPQR}\,
F^{(0)}_{LPQR}\,.\label{dP2}
\end{eqnarray}
Let us now turn to considering the equation of motion for the $C_{IJK}$ fields. They read:
\begin{eqnarray}
\partial_\mu\left(V_7\,\sqrt{-g} G^{I_1 J_1}G^{I_2 J_2}G^{I_3 J_3}\,
g^{\mu\nu}\, \partial_\nu C_{J_1J_2J_3}\right)&=&-\frac{3}{2}\,\frac{1}{7!}\,\epsilon^{\mu\nu\rho\lambda}\, F_{IJKP}\, F_{\mu\nu\rho\lambda}\, \epsilon^{I_1I_2I_3 IJKP}+\nonumber\\&&-\frac{1}{2}\,V_7\,\sqrt{-g}\,
\tau_{PQ}{}^{[I_1}\, F^{I_2I_3]PQ}\,.\label{dC}
\end{eqnarray}
By using equations (\ref{dP1}) and (\ref{dP2}) and the fact that:
\begin{eqnarray}
\frac{\delta ( F^{(0)}_{IJKL}\,F^{(0)\,IJKL})}{\delta C_{PQR}}&=&-3\,
\tau_{IJ}^{[P}\, F^{(0)QR]IJ}\,,\label{delta2}
\end{eqnarray}
equation (\ref{dC}) can be rewritten in the form:
\begin{eqnarray}
\partial_\mu\left(V_7\,\sqrt{-g} G^{I_1 J_1}G^{I_2 J_2}G^{I_3 J_3}\,
g^{\mu\nu}\, \partial_\nu C_{J_1J_2J_3}\right)&=&\sqrt{-g}\,\frac{\delta V}{\delta C_{I_1I_2I_3}}\,,
\end{eqnarray}
where the $C_{IJK}$--dependent part of the potential is:
\begin{eqnarray}
V_C &=& \frac{3}{16}\,\frac{1}{7!}\,\frac{1}{V_7}\, \left( C_{IJK}\,(g_{LPQR}+\frac{3}{4}\,\tau_{[LP}^N\,C_{QR]N})\,\epsilon^{IJKLPQR}+\tilde{g}\right)^2+\nonumber\\
&&+\frac{1}{6}\,V_7\,F^{(0)}_{IJKL}\,F^{(0)}_{MNPQ}\, G^{IM}\, G^{JN}\, G^{KP}\, G^{LQ}\,,\label{vc}
\end{eqnarray}
where $F^{(0)}_{IJKL}$ is given in eq. (\ref{eq15}). One can
easily compute the scalar potential in the Einstein frame by
noting that
\begin{eqnarray}
g_{\mu\nu}&=&\frac{1}{V_7^2}\, g_{\mu\nu}^{E}\,.
\end{eqnarray}
Therefore in this frame, the potential becomes multiplied by an overall $(V_7)^{-2}$.\par
The full scalar potential in the Einstein frame is thus obtained by adding to
$V_C$ the Scherk--Schwarz purely $G$--dependent part originating from the eleven--dimensional Einstein term. It is useful to write the entire potential as the following sum:
\begin{eqnarray}
V&=&V_E+V_K+V_{C-S}\,,\label{v}
\end{eqnarray}
 where the three terms on the right hand side originate from the eleven dimensional Einsten, kinetic and Chern--Simons terms respectively, and are found to have the following expression:
\begin{eqnarray}
V_E&=&\frac{1}{V_7}\,\left(2\, G^{KL}\, \tau_{KJ}{}^I\, \tau_{LI}{}^J+G_{II'}\,G^{JJ'}\, G^{KK'}\,\tau_{JK}{}^I\,\tau_{J'K'}{}^{I'}\right)\,,\nonumber\\
V_K&=&\frac{3}{16}\,\frac{1}{7!}\,\frac{1}{V_7}\,(g_{IJKL}+\frac{3}{2}\,\tau_{[IJ}^R\,C_{KL]R})(g_{MNPQ}+
\frac{3}{2}\,\tau_{[MN}^R\,C_{PQ]R})\, G^{IM}\, G^{JN}\, G^{KP}\, G^{LQ}\,,\nonumber\\
V_{C-S}&=& \frac{1}{6}\,\frac{1}{V_7^3}\, \left( C_{IJK}\,(g_{LPQR}+\frac{3}{4}\,\tau_{[LP}^N\,C_{QR]N})\,\epsilon^{IJKLPQR}+\tilde{g}\right)^2\,.\label{vekcs}
\end{eqnarray}
Note that for $\tau=g=0$ we just get a positive cosmological constant, as noted in \cite{d}.
 \section{ Flat group vacua of the potential}
The scalar potential in (\ref{v}) and (\ref{vekcs}) has the property that $V_K\ge 0$, $V_{C-S}\ge 0$
while $V_E$ has no definite sign \cite{ss}. Therefore in general we may have vacua with different signs of the cosmological constant.\par
A particular appealing class of models, which correspond to ``no--scale'' supergravities \cite{noscale},
 are obtained for those gaugings for which $V_E\ge 0$. This defines a ``flat group'' and implies restrictions
  of the $\tau$ matrices. These were described in the pioneering paper of ref. \cite{ss} for $g_{IJKL}=\tilde{g}=0$. \par
It is our aim to show here that for any flat group at $g_{IJKL}=\tilde{g}=0$ there is a new flat gauge
 algebra with additional structure constants related to $g$ and $\tilde{g}$. We first discuss this situation
 by looking at the flat vacua of our potential and, in the next section, from the point of view of the gauge structure. \par
To find extrema with zero cosmological constant of the scalar
potential we first analyze the equation $\delta {\Scr L}/\delta
C_{IJK}=0$. Because of the properties (\ref{dP2}) and
(\ref{delta2}) this is ensured by setting $F_{IJKL}=0$. The
contribution of the $C_{IJK}$ field to the  vacuum energy vanishes
if  $P=0$ at $F_{IJKL}=0$. This can be always obtained by a
particular choice of $\tilde{g}$ as a function of $g$. The
necessary conditions for a flat vacuum can be then summarized as
follows:
\begin{eqnarray}
F^{(0)}_{IJKL}&=&0\,\,\,\,\Leftrightarrow\,\,\,\,\,g_{IJKL}+\frac{3}{2}\, \tau_{[IJ}{}^P\, C_{KL]P}=0\,,\label{f40}\\
P&=&0\,\,\,\,\Leftrightarrow\,\,\,\,\,C_{MNR}\,\left(g_{IJKL}+\frac{3}{4}\, \tau_{[IJ}{}^P\, C_{KL]P}\right)\,\epsilon^{MNRIJKL}+\tilde{g}=0\,.\label{P0}
\end{eqnarray}
The second equation can also be written as the following condition on $\tilde{g}$:
\begin{eqnarray}
\tilde{g}&=&\frac{3}{4}\, C^0_{MNR}\,\tau_{[IJ}{}^P\, C^0_{KL]P}\,\epsilon^{MNRIJKL}\,,\label{tildeg}
\end{eqnarray}
where $C^0_{IJK}$ is a solution of equation (\ref{f40}) and thus depends on $g_{IJKL}$. This equation
ensures that the $G_{IJ}$ moduli equations are the same as in the $g=\tilde{g}=0$ case, because the
$F$--contribution to the energy--momentum tensor vanishes in these vacua.\par
To make a concrete example, let us consider the case in which $I=0,i$, $i=1,\dots, 6 $ with
$\tau_{IJ}{}^K=\tau_{0i}{}^j$, zero otherwise, and $g_{IJKL}=g_{0ijk}$, zero otherwise.
In this case $\tau_{0i}{}^j=T_j{}^i$ is chosen to be an antisymmetric matrix of rank 3 which can be set in the
form:
\begin{eqnarray}
T_i{}^j&=&\left(\matrix{m_1\,\epsilon &0&0\cr 0&m_2\,\epsilon &0\cr 0&0&m_3\, \epsilon}\right)\,\,\,;\,\,\,\,\epsilon=\left(\matrix{0&1\cr -1&0}\right)\,.
\end{eqnarray}
In this context  the equation (\ref{f40}) becomes
$F^{(0)}_{0ijk}=0$ which fixes all $C_{ijk}$ fields but not the
$C_{0ij}$ scalars. The $C_{0ij}$ fields give masses to the
$A_{ij}$ vector fields with the exception of the three entries
$(ij)=(1,2),(3,4),(5,6)$. Therefore three of the $C_{0ij}$ scalar
remain massless moduli. The $G_{IJ}$--sector gives, as discussed
in reference \cite{ss}, four additional massless scalars, of which
two are the volume $V_7$ and $G_{00}$ and two other come from
internal components of the metric.\par If one further discusses
the spectrum of the remaining fields, the six vector $A_{i0}$ are
eaten by the six antisymmetric tensors $B_i$ because of the
magnetic mass term in the FDA (\ref{eq1}). An additional massless
scalar comes from the massless 2--form $B_0$ and finally an
additional massless vector come from the $A^0$ Kaluza--Klein
vector. The other six $A^i$ vectors become massive because of the
twisting of the torus. We conclude that in this theory there are
always eight massless scalars and four massless vectors, in
agreement with \cite{ss}. The effect of turning on $g$ and
$\tilde{g}$ is not of giving extra masses, but of shifting the
v.e.v. of the $C_{IJK}$ fields. This can be understood by an
extension of the flat group where $g$ and $\tilde{g}$ play the
role of additional structure constants. In the next section we
will recover this result as well as the form of the potential,
from the underlying duality symmetry of the dual formulation of
the theory, in which all antisymmetric tensors $B_I$ are dualized
into scalars $\tilde{B}^I$ and the $\rE_{7(7)}$ symmetry is
recovered.
\section{The dual gauge algebra and its scalar potential}
We now interpret the above result in the usual formulation of the
four dimensional theory based on the flat gauging. From the
results of \cite{dft2} this amounts to dualize those vector fields
which participate to the anti--Higgs mechanism, in our case they
are the $A_{0i}$ 1--forms, which are therefore replaced by their
$A^{0i}$ magnetic duals. The dual gauge algebra therefore contains
the following 28 generaors:
\begin{eqnarray}
W^{ij},\, W_{i},\,Z_i,\,Z_0\,.
\end{eqnarray}
with structure constants obtained from eq. (2.13) of \cite{dft2}. The first 27 generators form
an abelian algebra, and the only non vanishing commutators are those involving $Z_0$ and given
by:
\begin{eqnarray}
\left[Z_0,\,Z_i\right]&=&T_j{}^j\,Z_j-12\,g_{0ijk}\, W^{jk}+\tilde{g}\, W_i\,\nonumber\\
\left[Z_0,\,W^{pq}\right]&=&2\, T_i{}^{[p}\,W^{q]i}-12\,g_{0ijk}\, \epsilon^{ijkpql}\, W_l\,\nonumber\\
\left[Z_0,\,W_i\right]&=& T_j{}^j\,W_j\,,\label{gaugealg}
\end{eqnarray}
where with respect to \cite{dft2} the redefinition $g\rightarrow
-12\,g$ was made. This algebra defines a flat subalgebra of
$\rE_{7(7)}$ which fits the class of models duscussed by Cremmer,
Scherk and Schwarz in \cite{css} and in \cite{svn}, as it was
shown in \cite{alt} and in \cite{dft2}. The gauged supergravity
interpretation was given in \cite{adfl} and the corresponding
gauge algebra is the semidirect product of a $\rU(1)$ by a
27--dimensional abelian algebra and is contained in the branching
of $\rE_{7(7)}$ with respect to $\rE_{6(6)}\times \rO(1,1)$:
\begin{eqnarray}
{\bf 133}&\rightarrow & {\bf 1}_0+{\bf 78}_0+{\bf 27'}_{+2}+{\bf 27}_{-2}\,.
\end{eqnarray}
To compare with the geometrical twist we further branch  $\rE_{6(6)}$ with respect to $\rSL(6)\times \rSL(2)$:
\begin{eqnarray}
{\bf 78}&\rightarrow & ({\bf 35},{\bf 1})+({\bf 1},{\bf 3})+({\bf 20},{\bf 2})\,,\label{branch78}\\
{\bf 27}&\rightarrow & ({\bf 15}',{\bf 1})+({\bf 6},{\bf 2})\,.
\end{eqnarray}
Our gauging corresponds to the following choice of the ``twist matrix'' (see \cite{alt} and equation (2.9) of \cite{dft2}):
\begin{eqnarray}
Z_0&=&-\frac{2}{3}\,T_i{}^j\, t_j{}^i+\,g_{0ijk}\,
t^{ijk}+\frac{1}{9}\,\tilde{g}\, t_0\,,
\end{eqnarray}
where we have used the notations introduced in \cite{dft2}. Here $t_i{}^j$ are the generators of the maximal compact subgroup of $\rSL(6)$, namely $\rSO(6)$, while $t^{ijk}$ and $t_0$ are nilpotent generators: the former belong to the $({\bf 20},{\bf 2})$ representation in (\ref{branch78}) with positive grading with respect to the $\mathfrak{o}(1,1)$ generator of $\rSL(2)$ and the latter is the nilpotent generator of  $\rSL(2)$ with positive grading with respect to the same generator. In the same framework we now discuss the form of the scalar potential, which is expected not to depend on the dualization procedure. In the dual formulation this potential is given by \cite{svn,adfl,dgftv}:
\begin{eqnarray}
V&=&e^{-6\,\phi}\,\left( \frac{1}{2}\,(P_{0\,\hat{i}}{}^{\hat{j}})+\frac{1}{6}\,(P_{0\,\hat{i}\hat{j}\hat{k}})^2+(P_0{}^0)^2\right)=V_E+V_K+V_{C-S}\,,
\end{eqnarray}
where  $\phi$ is the modulus associated with the $0^{th}$ internal
dimension of compactification, the hatted indices are rigid
$\rSO(6)$ indices, while the quantity $P_0$ has value  in the
42--dimensional non--compact part of the $\mathfrak{e}_{6(6)}$ Lie
algebra and represents the vielbein of the five--dimensional
$\sigma$--model $\rE_{6(6)}/\rUSp(8)$.  It is defined as follows:
\begin{eqnarray}
P_0=(L^{-1}\, Z_0\, L)_{|\mbox{non--compact}}\,,
\end{eqnarray}
where $L$ is the five--dimensional coset representative which, using the solvable Lie algebra parametrization of $\rE_{6(6)}/\rUSp(8)$, can be directly written in terms of our scalar fields as follows:
\begin{eqnarray}
L&=& e^{\tilde{B}^0\, t_0}\, e^{\frac{1}{6}\,C_{ijk}\, t^{ijk}}\,
\mathbb{E}\,\,\,;\,\,\,\,\mathbb{E}\in \frac{\rGL(6)}{\rSO(6)}\,.
\end{eqnarray}
Direct computation shows that:
\begin{eqnarray}
P_{0\,\hat{i}\hat{j}}&=&T_i{}^j\, \mathbb{E}^{-1}_{(\hat{i}}{}^i\, \mathbb{E}_{j|\hat{j})}\,,\nonumber\\
P_{0\,\hat{i}\hat{j}\hat{k}}&\propto &(g_{0ijk}+\frac{3}{4}\,T_{[i}{}^n\,C_{jk]n})\,\mathbb{E}^{-1}_{\hat{i}}{}^i\,\mathbb{E}^{-1}_{\hat{j}}{}^j\,\mathbb{E}^{-1}_{\hat{k}}{}^k\,,\nonumber\\
P_{0}{}^0&\propto & \epsilon^{lmnijk}\, C_{lmn}\,(g_{0ijk}+\frac{3}{8}\,T_{[i}{}^n\,C_{jk]n})+
\tilde{g} \,.\label{ps}
\end{eqnarray}
In this language the eight massless modes come from $\tilde{B}^0$, three from $C_{0ij}$, one from
$\phi$ and three from the metric $G_{ij}$. The latter can be understood from the fact that under $\rSO(6)$ these moduli transform in the ${\bf 1}+{\bf 20}^\prime$ and the ${\bf 20}^\prime$ has two vanishing weights.
\section{Conclusions and outlook}
In the present investigation we have presented the full set of (bosonic) curvatures and their
Bianchi identities for the free differential algebra underlying M--theory compactified on a
twisted seven--torus with the 4--form flux turned on. The resulting curvatures show that the
$B_I$ 2--forms receive a mass from the 1--form fields $A_{IJ}$ through the twist matrix
$\tau$, see equation (\ref{eq0}).
 Moreover the $B_I$ curvature $H_I$ shows that a Green--Schwarz coupling \cite{gs} is
 present which modifies the corresponding  Bianchi identity (\ref{eq11}).
In deriving the results of sections 2,3, and 4, we have used an
expansion of the eleven dimensional 3-form and of its 4-form
curvature in terms of the internal twisted torus with fluxes,
along similar lines as those discussed in references \cite{km,df}.
By projecting the FDA and the equations of motion on a suitable
basis of 1-forms one
recovers the main formulae of sections 2,3, and 4.\\
 Combined gauge invariance of tensors of different rank has been also considered in a different framework in
\cite{dws}. We have derived the scalar potential (eqs. (\ref{v}), (\ref{vekcs})) and discussed
the flat group vacua giving rise to a no--scale form of $N=8$ supergravity \cite {adfl}.\par
   Moreover we have compared the results with a dual formulation of the theory
   where those vectors participating to the anti--Higgs phenomenon have been dualized together with the seven antisymmetric tensors
so that the $\rE_{7(7)}$ symmetry is restored. In the dual formulation the scalar potential is
given in the form obtained by dimensional reduction from five dimensions \cite{svn,adfl} in
the presence of a suitable $\rE_{6(6)}$--twist. It is proven that the two potentials do
coincide, giving evidence that the theories are ``dual'' even in the presence of gaugings.
This is however possible only for a restricted gauge algebra where the $\rU(1)$ compact
generator $Z_0$ is taken in the (rank three) maximal compact subgroup $Spin(6)=\rSU(4)$ of
$\rSL(4)$, rather than in the rank--4 maximal compact subgroup $\rUSp(8)$ of $\rE_{6(6)}$. The
missing compact generator lies inside the $\mathfrak{sl}(2)$ subalgebra of
$\mathfrak{e}_{6(6)}$ commuting with $\mathfrak{sl}(6)$. It would be interesting to compute
the full Lagrangian reduced to $D=4$ and its dual theory with the underlying gauge algebra of
reference \cite{dft2}. This would involve dualization of massive tensors along the same lines
of references \cite{ddsv}-\cite{s}

\par We have not discussed vacua with $F^{(0)}_{IJKL}$ and/or $P$ not
vanishing. First of all we may notice that , on general grounds, if we assume
$G_{IJ}=\lambda^2\,\delta_{IJ}$, the three terms in (\ref{vekcs}) scale differently; for large
$\lambda$  they behave as $\lambda^{-9}, \lambda^{-15}, \lambda^{-21}$, respectively, so the
sign of $V_{E}$ dominate the potential for large $\lambda $.\\
The Einstein equations, (\ref{einst}), may have solutions where the gravitational part is
compensated by the energy momentum tensor of the $C_{IJK}$ scalars. The equation of motion,
given in (\ref{dC}), may have a solution for $F^{(0)}_{IJKL}\neq 0$. We note that these
equations are homogeneous in $F^{(0)}_{IJKL}$. The solutions with non flat vacua may be
important for cosmological applications.\\
 We also note that in the FDA formulation of the theory
the R--symmetry is $Spin(7)$ while in the dual formulation is
  $Spin(6)=\rSU(4)$. The eight gravitinos belong to the ${\bf 8}$ of $Spin(7)$ in one formulation
and to the ${\bf 4}+\overline{{\bf 4}}$ of $Spin(6)$ in the dual formulation.
When the fermionic sector  is included in the theory, the bosonic FDA extends to a ``super'' FDA,
that we call SFDA. The curvatures and Bianchi identities of the full theory will then include
 the spin--$3/2$ curvatures and possibly the spin--$1/2$ curvatures. Their construction
 is under investigation and will be given elsewhere.\par
Finally it is possible to compute several truncations of this $N=8$ theory in order to obtain
lower ($N<8$) local supersymmetry.\par
The scalar potential of the reduced theory will be obtained in this case by a particular
 truncation of the original potential \cite{dgftv}. \par
It should be emphasized that there is a subtlety in the computation of the M--theory potential,
which is the term $V_{C-S}$ of equation (\ref{vekcs}).
 This lies in the derivation of the equation of motion of the auxiliary field $P$ in (\ref{P}).
 In fact the contribution from the Chern--Simons
term does not come by merely treating $P$ as an algebraic Lagrange multiplier, but rather by
varying the potential ${\mathcal A}_{\mu\nu\rho}$ and solving a differential equation for $P$,
(\ref{dP}).
 Its integration yields equation (\ref{dP1}), which is in agreement with the result of ref. \cite{kachru},
  where a different derivation was used. It also gives an expression which vanishes at $F^{(0)}_{IJKL}=0$ by
 suitably adjusting the $\tilde{g}$ coupling, equation (\ref{tildeg}). This result is also in agreement with
the dual formulation (equation (\ref{ps})), where $\tilde{g}$ plays the role of a structure
constant of the gauge Lie algebra given in (\ref{gaugealg}).
\section{Acknowledgements}
We thank G. Dall'Agata for several discussions in the early stages of this investigation. \par
Work supported in part by the European Community's Human Potential Program under contract
MRTN-CT-2004-005104 `Constituents, fundamental forces and symmetries of the universe', in
which R. D'A. and M.T.  are associated to Torino University. The work of S.F. has been
supported in part by European Community's Human Potential Program under contract
MRTN-CT-2004-005104 `Constituents, fundamental forces and symmetries of the universe', in
association with INFN Frascati National Laboratories and by D.O.E. grant DE-FG03-91ER40662,
Task C.

\end{document}